\documentclass[]{aa}  

\usepackage{comment}
\usepackage{graphicx}
\usepackage{txfonts}
\usepackage{siunitx}
\usepackage{multirow}
\usepackage{amsmath, amssymb}
\usepackage{subfig}
\usepackage{rotating}
\usepackage{appendix}
\usepackage[dvipsnames]{xcolor}

\usepackage[
    starfontserif 
    ]{starfont}
\usepackage{amsmath}

\DeclareSymbolFont{starfontsym}{OT1}{sts}{m}{n}
\DeclareMathSymbol{\mathTerra}{\mathord}{starfontsym}{76}

\newcolumntype{Q}{>{\tiny}c}
\begin{document}

   \title{
    CO gaps in cold transition disks around low-mass stars}
   \subtitle{}

   \author{ A. A. Rota
          \inst{\ref{leiden}},
          N. van der Marel
          \inst{\ref{leiden}},
          E. F. van Dishoeck\inst{\ref{leiden}}, 
          T. Muto\inst{\ref{muto}}
        }

    \institute{Leiden Observatory, Leiden University, P.O. Box 9513, 2300 RA Leiden, The Netherlands\label{leiden}\\ \email{rota@strw.leidenuniv.nl}
    \and Division of Liberal Arts, Kogakuin University, 1-24-2 Nishi-Shinjuku, Shinjuku-ku, Tokyo 163-8677, Japan\label{muto}
    }
        
\authorrunning{Rota et al.}

   \date{Received -; accepted -}

\abstract{Transition disks with large inner dust cavities are promising targets to study planet formation, as these disks are likely hosting massive planetary companions. The structure of gas and dust inside the cavity are crucial for understanding the potential of planet formation in the regions.}{We investigate the structure of gas and dust in a sample of three cold, inclined transition disks around K and M stars without evidence for an inner dusty disk.}{We present new high-resolution ALMA Band 6 data of both 230 GHz continuum at 0$\farcs$02 ($\sim 3$ au) resolution and the $^{12}$CO 2--1, $^{13}$CO 2--1, C$^{18}$O 2--1 isotopologues at 0$\farcs$12 ($\sim 20$ au) resolution. We analyze the radial profiles, compute the brightness temperatures, and derive estimates of the gas and dust content of the inner disk at 1 au.}{All three disks show dust cavities of 30--40 au in radius. The CO emission extends inside the cavity with progressively smaller CO cavity radii for the more abundant isotopologues. The $^{13}$CO brightness temperature profile is similar to that of other transition disks and full disks. No inner dust disk is detected in the ALMA continuum, consistent with the lack of near-infrared excess. Considering the accretion rates, the gas-to-dust ratio in the inner disk at 1 au is estimated to be as high as 10$^5$.}{No significant differences in the brightness temperature or in the gas-to-dust ratio are found between our disks and other transition disks around more luminous stars, contrary to expectations. This highlights the importance of individual-target studies through thermochemical modeling or multi-transition analyses to constrain the dust and gas structures in the cavity. The observed disk structures could be explained by super-Jovian companion on eccentric orbits, but further modeling is needed to confirm this conclusion.}

\keywords{protoplanetary disks, stars: formation, (stars:) planetary systems}

\maketitle

\section{Introduction}\label{intro}

Transition disks are a class of protoplanetary disks with large cavities in their dust distribution (e.g., \citealp{2014Espaillat}, \citealp{2023VDMarelTDs}).  The commonly proposed scenario is that massive companions, such as giant planets or low-mass brown dwarfs, dynamically clear gaps in the disk and push material outward, creating a pressure maximum at the gap edge that traps drifting mm-sized dust particles (e.g., \citealp{2012Pinilla}). The result is a sharp, bright ring of millimeter dust just outside the orbit of the companion (e.g., \citealp{2014Perez}; \citealp{2015vdMarel}; \citealp{2016vanderMarel}). 
This hypothesis is supported by the direct detection of two actively accreting proto-planets in the cavity of PDS 70 (\citealp{2018Keppler}; \citealp{2021Benisty}).

Similar to the dust, the gas component also exhibits cavities, where the gas surface density drops by orders of magnitude. 
Modeling of CO isotopologue emission typically indicates a drop in the gas surface density by a factor of $\sim 10^2-10^4$ inside the cavity compared to the outer disk (e.g., \citealp{2014Bruderer}; \citealp{2016vanderMarel}; \citealp{2017ADong}; \citealp{2022Leemker}). Moreover, these studies have generally found that the gas cavity radius is smaller than the dust cavity radius, with the ring traced by $^{12}$CO,  $^{13}$CO, and  C$^{18}$O showing progressively larger radii according to their abundances and optical depths (e.g., \citealp{2018vanderMarel} \citealp{2021vdMarel}; \citealp{2023Woelfer}).

Therefore, material must continue to flow through the gap towards the central star.  High accretion rates onto the central star of $\sim 10^{-10} - 10^{-7}$ M$_\odot \text{yr}^{-1}$ (e.g., \citealp{2007Najita}; \citealp{2014Manara}), comparable to disks without large inner cavities, are observed. However, the way material filters through the gap and accretes onto the star is still unclear.
 Studying the gas and dust profile is thus crucial both to understand the flow of material in the cavity and to infer the properties of the possible embedded companion.

In this work, we investigate the gas and dust structure of three transition disks: Sz 111 (spectral type SpT M1.0), 2MASS J16335560-244204 (SpT K5.0), and  2MASS J16083070-3828268 (SpT K2.0). Unlike many prior studies, which focused on disks around hotter and more luminous stars, our sample consists of inclined transition disks around late-type (K/M), low luminosity ($\sim 1-2$ L$_\odot$) stars. We use ALMA Band 6 (1.3 mm) observations of the dust continuum and CO $J$=2–1 isotopologues ($^{12}$CO,  $^{13}$CO, and  C$^{18}$O) at $\sim0\farcs02$ and $\sim0\farcs1$ resolution, respectively. 

In Section \ref{sec:sampleanddatared}, we present the sample properties and the data analyzed in this work. In Section \ref{sec:resP4} and \ref{discussionp4}, we present and discuss the results. In Section \ref{discussionp4}, we draw our conclusions.

\begin{table*}[htbp]
\centering
\small
\caption{The sample of this work.}
\label{tab:sample}
\begin{tabular}{cccccccccc}
\hline\hline 
Target  & Coordinates & $d$ & SpT & $M_*$  & $L_*$ &  log $\dot{M}_\text{acc}$  & inc & PA (N-E) & Ref. \\
 & ICRS $J2000$ & [pc] & & [M$_\odot$] & [L$_\odot$] & [M$_\odot$ yr$^{-1}$] & [deg] &[deg] & \\
\hline 

Sz 111 & 16:08:54.684938, -39:37:43.145070 & 158.4 & M1.0 & $0.52$ & $0.21$ & $-9.4$ & 40 & $53$ & 1,2,5 \\

J1633 & 16:33:55.614857, -24:42:05.002391  & 143.5 & K5.0 & $0.80$ & $1.0$ & $-10.0$ & 77 & 48 & 4 \\

J1608 & 16:08:30.699406, -38:28:26.846324 & 153.4 & K2.0 & $1.27$ & $1.9$ & $-9.2$ & 107 & $74$ & 1, 3, 5 \\

\hline
\end{tabular}
\tablefoot{Distances from Gaia DR3 \citep{gaiaDR3}.
References: 1) \cite{2018vanderMarel}; 2) \cite{2014Alcala}; 3) \cite{2017Alcala}; 4) \cite{2021Cieza}; 5) \cite{2023ManaraPPVII}
}
\end{table*}

\section{Sample and data reduction}\label{sec:sampleanddatared}

\subsection{Sample properties}

The sample consists of three targets: two systems are part of the Lupus star-forming region -- Sz 111 and 2MASS J16083070-3828268 (hereafter J1608) -- and one part of Ophiuchus -- 2MASS J16335560-2442049 (hereafter J1633), located at distances between 140 and 160 pc.

Table \ref{tab:sample} summarizes the properties of the three disks and their stars, after correcting for the Gaia DR3 distances \citep{2023ManaraPPVII}. The position angle (measured east-of-north) and inclination of the disks have been taken from the literature based on analysis of the continuum emission at lower resolution \citep{2018vanderMarel,2021Cieza}.
The stars in these systems are late-type stars (spectral types M and K), with luminosity ranging from $0.2$ L$_\odot$ to $2$ L$_\odot$ and stellar masses from $0.5$ M$_\odot$ to $1.3$ M$_\odot$. The stars are associated with inclined ($i>40\deg$) transition disks with dust cavities of $\sim 30 - 40~$ au in radius (e.g., \citealp{2016Ansdell, 2018Ansdell,2018vanderMarel}; \citealp{2021Cieza}). The accretion rates onto the central star are $\sim 10^{-10} - 10^{-9} M_\odot~$yr$^{-1}$.

\subsection{Observations and data reduction}

The disks analyzed in this work were observed in ALMA Band 6 (program 2018.1.00689.S, PI: Muto) in both long-baseline (integration time $\sim 50$ min per target, angular resolution $\sim0''.019$, continuum sensitivity $\sim0.012$ mJy beam $^{-1}$) and short-baseline (integration time $\sim 10$ min per target, angular resolution $\sim0''.109$, continuum sensitivity $\sim0.030$ mJy beam$^{-1}$) configurations.
The spectral setup included two continuum spectral windows centered at 216 and 232 GHz, respectively and three spectral windows targeting the molecular line emission of $^{12}$CO ($J$=2-1), $^{13}$CO ($J$=2-1), and C$^{18}$O ($J$=2-1) at 488 kHz spectral resolution.

The calibrated data were retrieved using the script for PI provided by the observatory. 
To improve the signal-to-noise ratio (SNR), we attempted self-calibration; however, since the resulting improvement was $<10\%$, self-calibration was not applied to the final dataset analyzed here.
Continuum and line images were created with the CASA task \texttt{tclean}. Continuum images were made using both the continuum spectral window and the line-free channels of the three spws in which CO isotopologue emission was detected. The cleaning was performed with Natural weighting using an interactive cleaning. For the CO isotopologues, the cleaning was performed with a channel width of $1.0$ km s$^{-1}$. Only the short-baseline data were included in the imaging of the CO isotopologues, due to low signal-to-noise of the highest resolution data (resulting SNR for $^{12}$CO $<5$). The typical beam size of the continuum emission is $0''.03$ or 30 mas, whereas the beam of the imaged line emission is typically $0''.13$ or 130 mas.
Table \ref{tab:rms_beam_cont} reports the rms and beam size of the continuum image and of the zeroth moment maps of the CO isotopologues. Figure \ref{fig:mapsCOCont} presents the continuum and the zeroth moment maps of the CO isotopologues, while Figure \ref{fig:mapsVelocity} shows the spectra and the velocity maps of the three targets.

\begin{table}[]\centering\small
\caption{Information on the cleaned images}
        \begin{tabular}{cccc}

        \hline\hline
                Target   & & rms & Beam size\\
                   & & [mJy beam$^{-1}$ per channel] & [$\arcsec \times \arcsec$] \\

                   \hline

                   Sz 111   & \multirow{3}{*}{}continuum  &        1.43e-02                &   $ 0.035 \times 0.028$  \\
                                                &$^{12}$CO  &   2.36               &    $0.138 \times 0.113$    \\
                                                &$^{13}$CO  &   2.45                &    $0.144 \times 0.119$   \\
                                                &C$^{18}$O  &   1.97                 &    $0.145 \times 0.120$    \\

                \hline
                J1633 & \multirow{3}{*}{}continuum  & 1.52e-02      &    $0.037 \times 0.031$   \\
                                                &$^{12}$CO  &   2.25                 &    $0.127 \times 0.110$    \\
                                                &$^{13}$CO  &   2.39                &    $0.132 \times 0.115$    \\
                                                &C$^{18}$O  &   1.94                &    $0.133 \times 0.116$   \\
            \hline
            J1608   & \multirow{3}{*}{}continuum  &    1.30e-02                &    $0.027 \times 0.024$    \\
                                                &$^{12}$CO  &    2.33                &    $0.143 \times 0.112$     \\
                                                &$^{13}$CO  &   2.47                &    $0.149 \times 0.117$   \\
                                                &C$^{18}$O  &   2.02           &    $0.151 \times 0.118$    \\
            \hline
                
        \end{tabular}
\label{tab:rms_beam_cont}
\end{table}

\section{Results}\label{sec:resP4}

\begin{table}[]\centering\small
\caption{Disk fluxes and radii}
        \begin{tabular}{ccccc}

        \hline\hline
                Target   & & Total flux & Disk radius & Peak\\
                   & & [mJy or mJy km/s] & [au] & [au] \\

                   \hline
                Sz 111   & \multirow{3}{*}{}continuum  &       $83.3 \pm  3.1$      &    $114.5 \pm 2.8$ &  39 \\
                                                &$^{12}$CO  &   $3104\pm264$              &    $254\pm13$ & 17   \\
                                                &$^{13}$CO  &   $1067\pm206$               &    $224^{+13}_{-24}$ & 24   \\
                                                &C$^{18}$O  &   $254\pm155$                &    $73^{+98}_{-24}$ & 31  \\
            
\hline
                J1633 & \multirow{3}{*}{}continuum  & $100.4 \pm 0.3$                &    $62.6 \pm 2.7$  &  35 \\
                                                &$^{12}$CO  &   $4348\pm260$                 &    $206\pm11$ & 3   \\
                                                &$^{13}$CO  &   $1064\pm220$                &    $133^{+20}_{-25}$ & 21  \\
                                                &C$^{18}$O  &   $540\pm164$              &    $86^{+37}_{-26}$  & 27 \\
            \hline
            J1608   & \multirow{3}{*}{}continuum  &   $124.7 \pm 0.5$                &    $213.5^{+2.5}_{-3.3}$  & 40 \\
                                                &$^{12}$CO  &   $18271\pm736$               &    $307\pm11$  & 29   \\
                                                &$^{13}$CO  &   $7443\pm397$                &    $301\pm11$  & 30 \\
                                                &C$^{18}$O  &   $1559\pm187$          &    $\sim 330^*$ & 64   \\

                \hline
        \end{tabular}
        \tablefoot{For each system and line, we indicate the detected total flux and disk radius, estimated with the cumulative flux technique. The last column reports the location of the peak of the CO and continuum emission. $^*$ The estimate of the C$^{18}$O outer radius is uncertain due to low SNR.}
\label{tab:fluxes_radii}
\end{table}

\subsection{Dust and gas structure}

\begin{figure*}
   \centering
   \includegraphics[width=\textwidth]{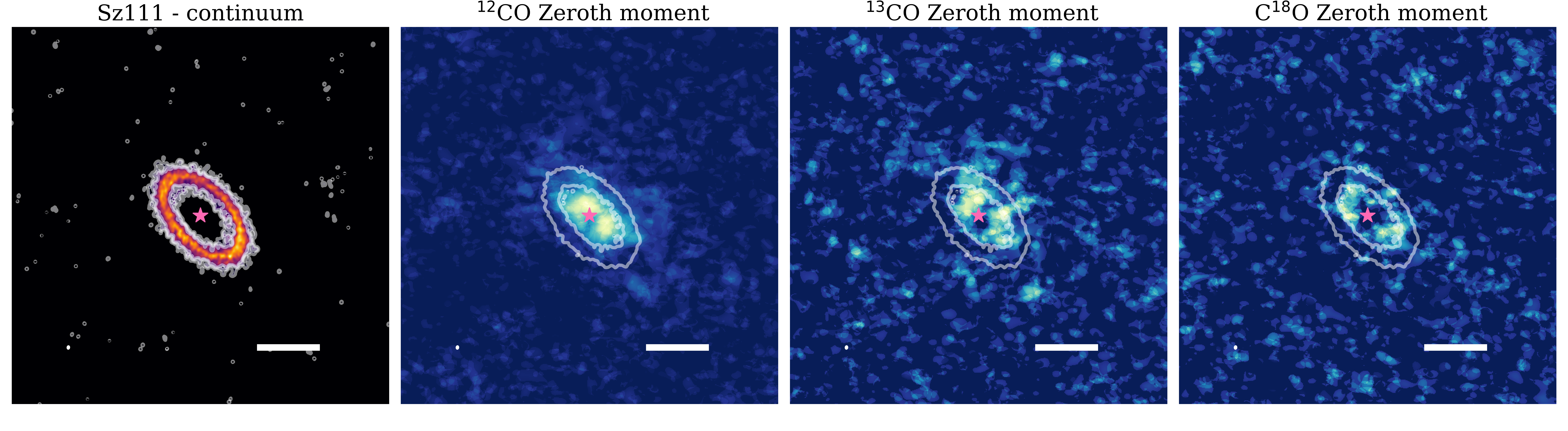}
    \includegraphics[width=\textwidth]{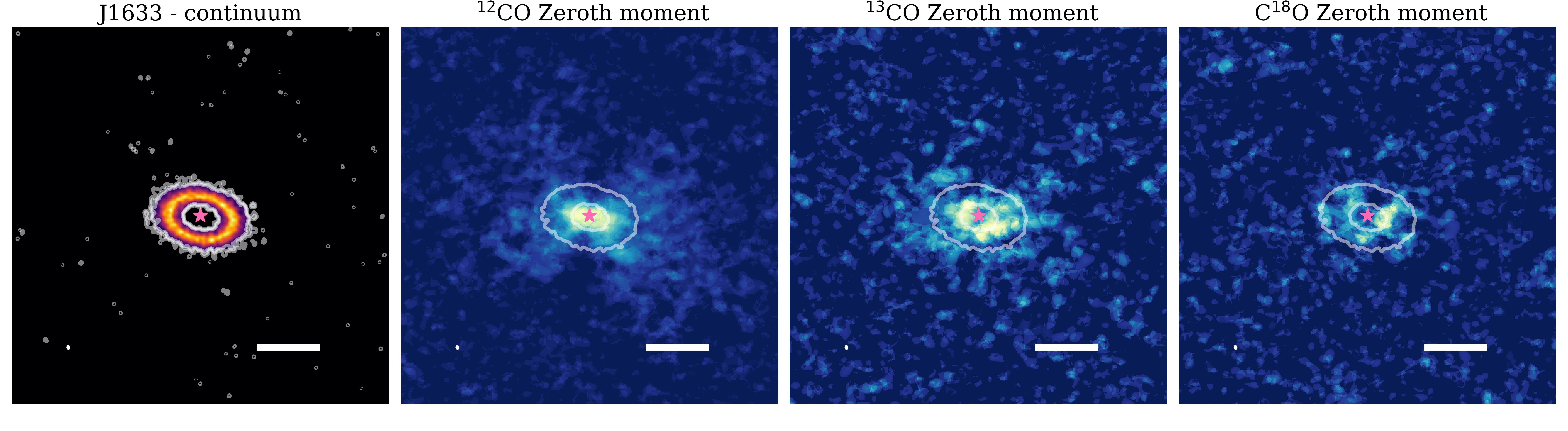}
    \includegraphics[width=\textwidth]{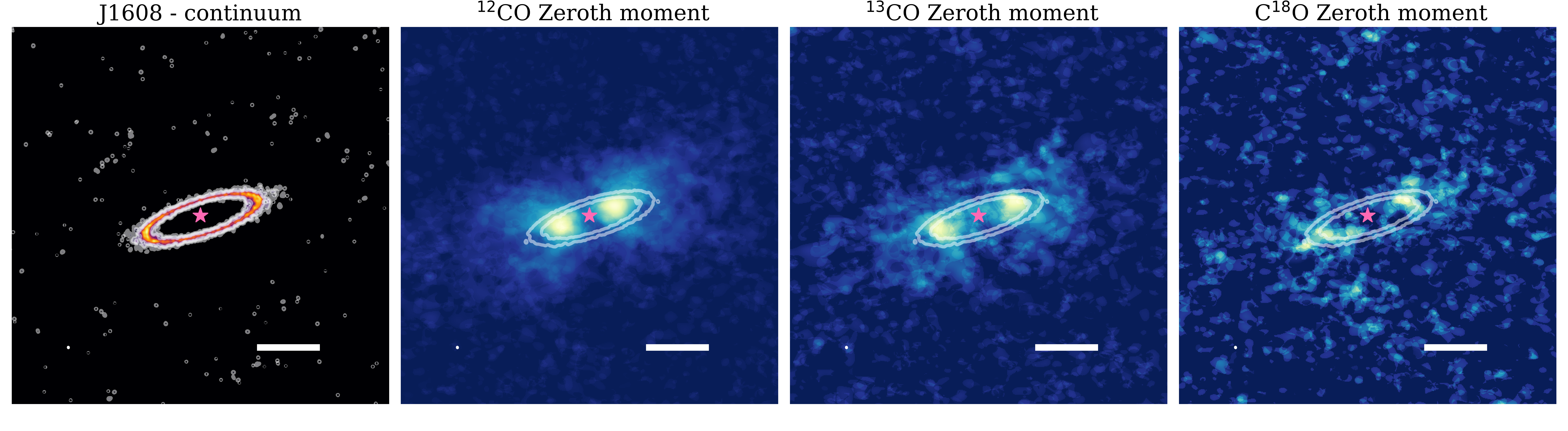}

      \caption{Continuum maps and zeroth moment maps of the CO isotopologues observed in ALMA Band 6. Each row refers to a target; the columns show the continuum, $^{12}$CO ($J$=2-1), $^{13}$CO ($J$=2-1), and C$^{18}$O ($J$=2-1) emission, respectively. Each panel is $3''.0 \times 3''.0$, which is $\sim(450\times450)$ au at a typical distance of 150 pc. All bars in the bottom right of each panel are $0''.5$ long ($\sim 75$ au). The beam is shown in the bottom left of each panel. In each image, the color scale has the peak flux of the emission as the maximum, and three times the image rms as the minimum. In the first column, white contours show 3, 5, and 7 times the rms of the continuum image; in all other columns, they show the dust emission at 7 times the rms of the continuum image. The star position is shown with a pink star. The moment zeroth maps were created with \texttt{bettermoments} \citep{2018TeagueForemanMackey} and a 2-sigma clipping process.}
    \label{fig:mapsCOCont}  
\end{figure*}

\begin{figure*}
   \centering
   \includegraphics[width=\textwidth]{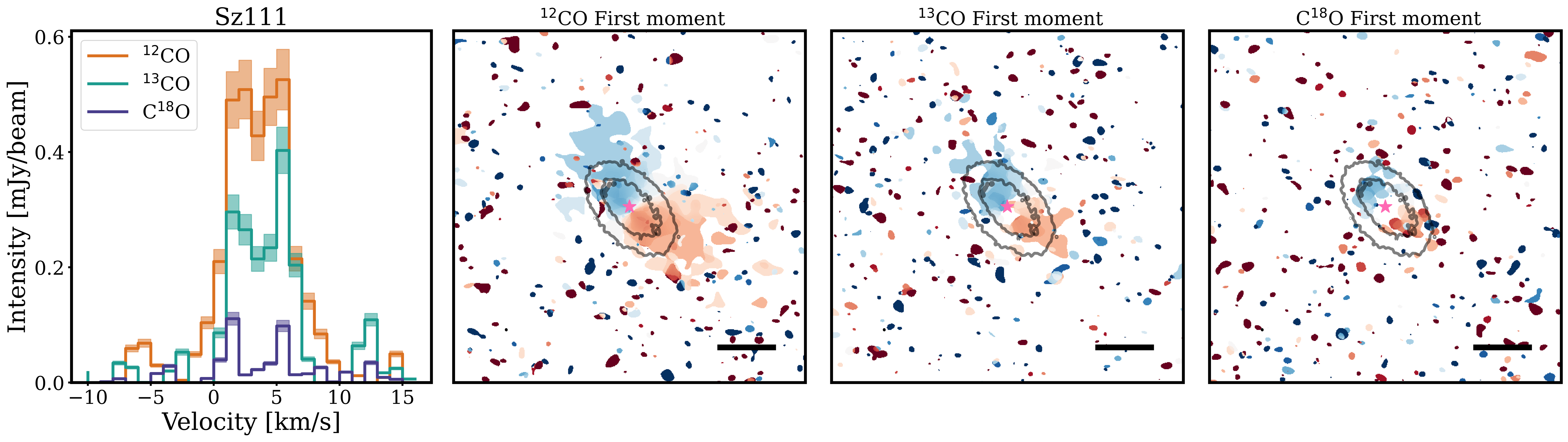}
    \includegraphics[width=\textwidth]{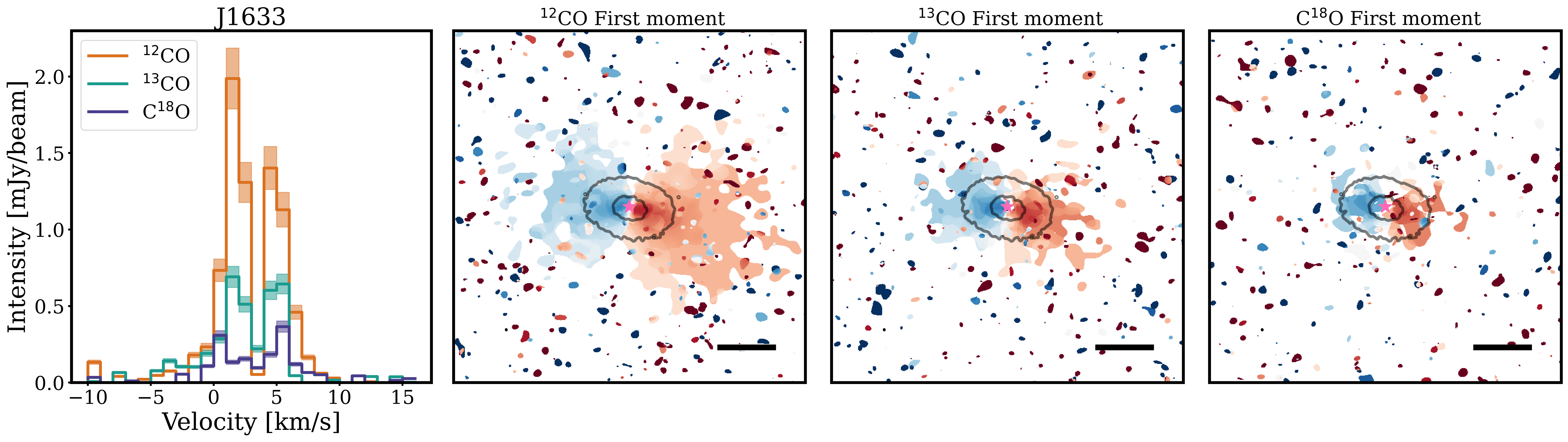}
    \includegraphics[width=\textwidth]{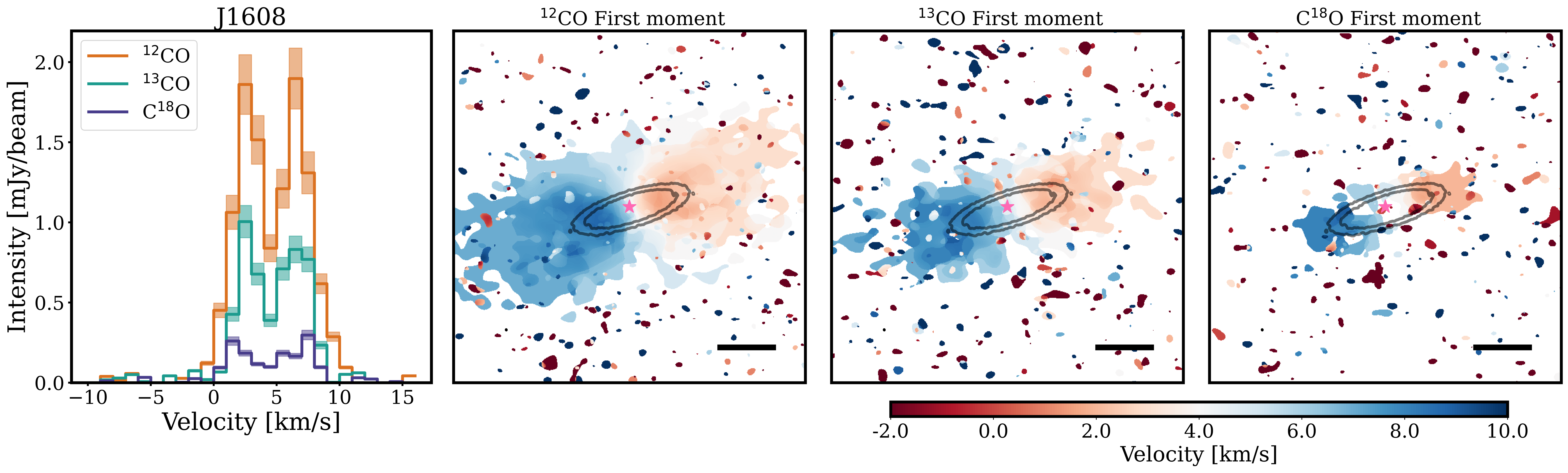}

      \caption{Spectra and first moment maps for the CO isotopologues. Each row refers to a target; the columns show the spectra of the lines, the $^{12}$CO ($J$=2-1), $^{13}$CO ($J$=2-1), and C$^{18}$O ($J$=2-1) maps, respectively. Each panel is $3''.0 \times 3''.0$, which is $\sim(450\times450)$ au at a typical distance of 150 pc. All bars in the bottom right of each panel are $0''.5$ long ($\sim 75$ au). The beam is shown in the bottom left of each panel. In each image, the color scale has the peak flux of the emission as the maximum, and three times the image rms as the minimum. The black contours show the dust emission at 7 times the rms of the continuum image. The star position is shown with a pink star. The first moment maps were created with \texttt{bettermoments} \citep{2018TeagueForemanMackey} and a 3-sigma clipping process.}
    \label{fig:mapsVelocity}  
\end{figure*}

\begin{figure*}
   \centering
   \includegraphics[width=\textwidth]{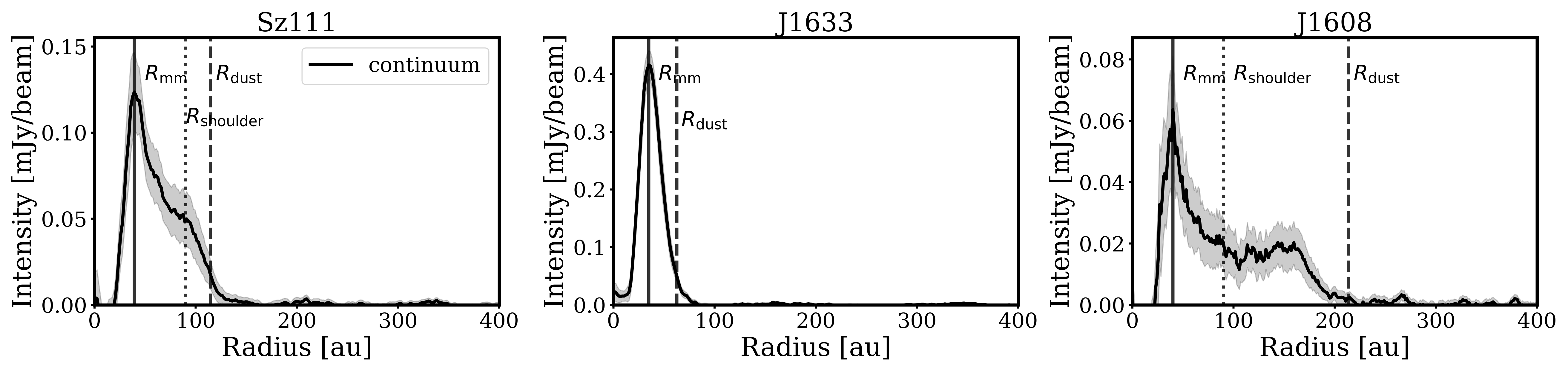}
    \includegraphics[width=\textwidth]{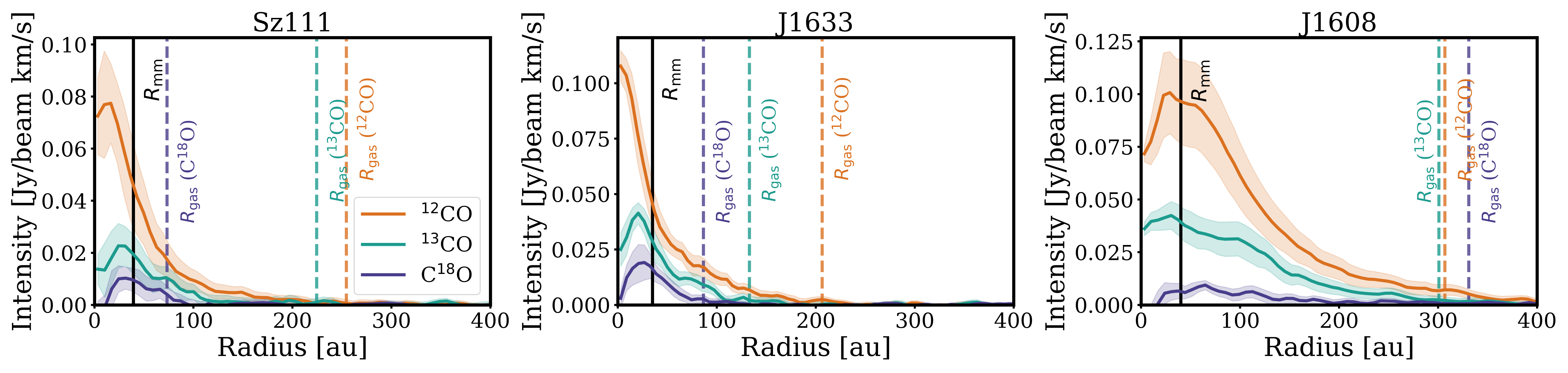}

      \caption{Deprojected azimuthally averaged radial profiles for the Band 6 continuum emission (\textit{top}) and for the CO isotopologue emission (\textit{bottom}), $^{12}$CO ($J$=2-1, orange), $^{13}$CO ($J$=2-1, green), and C$^{18}$O ($J$=2-1, purple). The solid black lines indicate the peak position of the dust ring ($R_\text{mm}$) in each disk. For Sz 111 and J1608, the dotted black lines indicate the shoulder ($R_\text{shoulder}$). The dashed black and colored lines show the estimated dust and gas radius of the disk, defined as the radius containing 95\% of the total flux of the continuum and each of the CO lines.}
    \label{fig:Radprof_cont}  
\end{figure*}

The ALMA continuum and CO isotopologue zeroth and first moment images of the three disks are presented in Figure \ref{fig:mapsCOCont} and Figure \ref{fig:mapsVelocity}. The three disks are detected and spatially resolved in both continuum and CO emission. The azimuthally averaged radial profiles are extracted from the continuum and zeroth moment maps, taking into account the position angle and inclination of the disks (Table \ref{tab:sample}) and are reported in Figure \ref{fig:Radprof_cont}.

The continuum maps reveal central cavities surrounded by narrow, azimuthally symmetric dust rings. No inner disks are detected down to the rms level. 
This morphology is highlighted in the deprojected, azimuthally averaged radial profiles of the three disks (top panel in Figure \ref{fig:Radprof_cont}), showing that the dust rings are located in the three disks at a similar distance from the host star -- 39 au, 35 au, and 40 au for Sz 111, J1633, and J1608, respectively. We note that the cavity sizes for Sz 111 and J1608 are significantly smaller than reported in \citet{2018vanderMarel}, due to the assumed distance of 200 pc in that work. 

The zeroth moment maps of the CO isotopologues generally exhibit central cavities as well, except for the $^{12}$CO emission in J1633 (Figure \ref{fig:mapsCOCont}). Line emission is detected within the dust cavity, similar to what is seen in other transition disks \citep[e.g., see][for an overview]{2023VDMarelTDs}, as clearly shown by the radial profiles (bottom panel in Figure \ref{fig:Radprof_cont}). In particular, the peaks of the CO isotopologue radial profiles are in all cases located at $\lesssim 40~$au, well inside the dust cavities.
Table \ref{tab:fluxes_radii} presents some properties of the disks, estimated from the radial profiles. The last column in Table \ref{tab:fluxes_radii} reports the peak location for each CO isotopologue. The $^{12}$CO emission peaks closer to the star than both the $^{13}$CO and C$^{18}$O, likely reflecting the effect of different optical depths of CO isotopologues according to their different abundances (e.g.,  \citealp{2017ADong}; \citealp{2022Leemker}). Overall, the CO isotopologue profiles are consistent with a ring-like morphology, except for the $^{12}$CO in J1633, which peaks at the stellar position.

A pattern compatible with Keplerian rotation is detected in the CO isotopologues. As shown by the spectra (Figure \ref{fig:mapsVelocity}), no strong signatures of foreground absorption by cloud material are found, although weak partial absorption around the system velocity is observed in all disks. 

We estimate the total flux and disk radii, in both CO and continuum emission, of each disk using an image plane analysis with the cumulative flux technique (e.g., \citealp{2018Ansdell}). For each target, the total flux is calculated by summing over concentric annuli centered on the star position (from Gaia DR3, \citealp{gaiaDR3}) and corrected for the projected position of the disk. The disk radius is then defined as the radius containing $95\%$ of the total flux (Figure \ref{fig:cumulatives}). The uncertainty on the radius accounts for both the error on the total flux and the resolution of the observations. Table \ref{tab:fluxes_radii} reports the total fluxes and radii estimated for each disk and line.

Among the three disks, the dust ring around J1633 is the narrowest, with an outer disk size of $R_\text{dust} = 62.7 \pm 2.7$ au, and an inner dust cavity radius of 35 au. In contrast, the continuum emission in Sz 111 and J1608 with dust cavity radii of $\sim$40 au extends to larger radii, $\sim 100-200$ au, and their radial profiles present a shoulder at $\sim 90~$ au, suggesting the possible presence of a second unresolved gap. The gaseous disks extend far beyond the dust disk radius, with an average $R_\text{gas}/R_\text{dust}$ (with $R_\text{gas}$ derived from the $^{12}$CO emission) of $\sim 2.3$, consistent with the average found in the complete sample of all detected protoplanetary disks in the Lupus star forming region (\citealp{2018Ansdell}). Overall, the radial structures and sizes of these disks are consistent with those of transition disks characterized in previous observational studies (see \citealp{2023VDMarelTDs} for a review).

\subsection{Brightness temperature of the CO isotopologues}

The brightness temperature profiles are derived using the peak flux of the CO lines (i.e., the eighth moment maps). We calculated the brightness temperature $T_b$ of the CO isotopologues both using the full Planck curve and the Rayleigh-Jeans approximation ($h \nu \ll k T$):

\begin{equation}
    T_b =
    \frac{h\nu}{k_B}
    \left[
    \ln\left(
    1 + \frac{2 h \nu^3}
    {c^2 \left(I_\nu/\Omega_\text{beam}\right)}
    \right)
    \right]^{-1}
    \qquad \text{Full Planck law,}
\end{equation}\begin{equation}
    T_b = \frac{c^2}{2 k_B v^2} \frac{I_\nu}{\Omega_\text{beam}} \qquad \text{Rayleigh-Jeans approximation,}
\end{equation}

where $c$ is the speed of light, $k_B$ is the Stefan-Boltzmann constant, $\nu$ is the observing frequency, and $I_\nu$ is the detected flux (in Jy beam$^{-1}$). $\Omega_\text{beam}$ is the solid angle, defined as the area of a Gaussian beam by its 2-dimension integral: $\Omega_\text{beam} = (\pi \theta_\text{maj} \theta_\text{min} ) / (4 \ln2) $, with $\theta_\text{maj}$ and $\theta_\text{min}$ the major and minor axis of the beam, respectively. 

\begin{figure*}
   \centering
   \includegraphics[width=\textwidth]{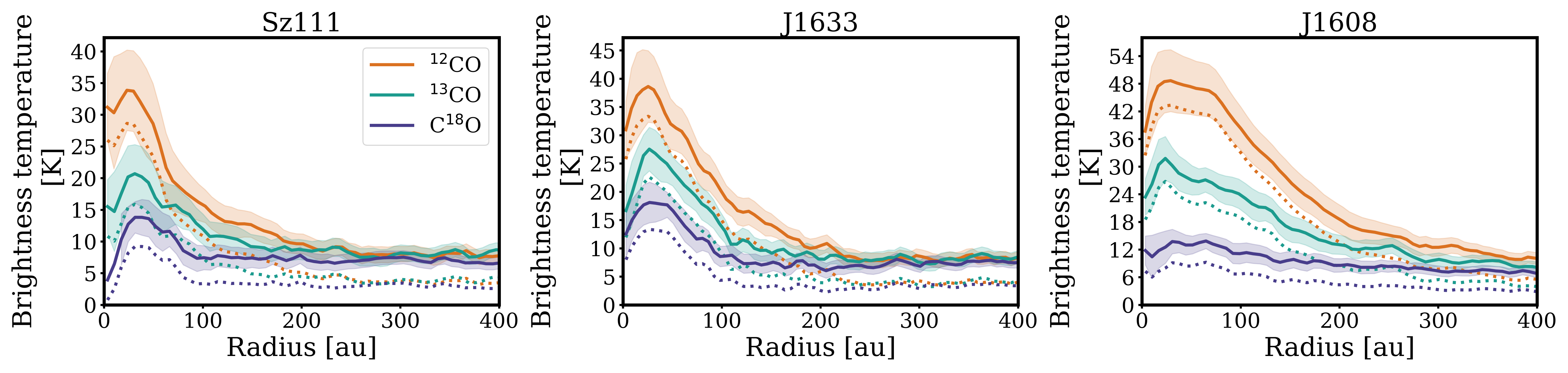}
   \includegraphics[width=\textwidth]{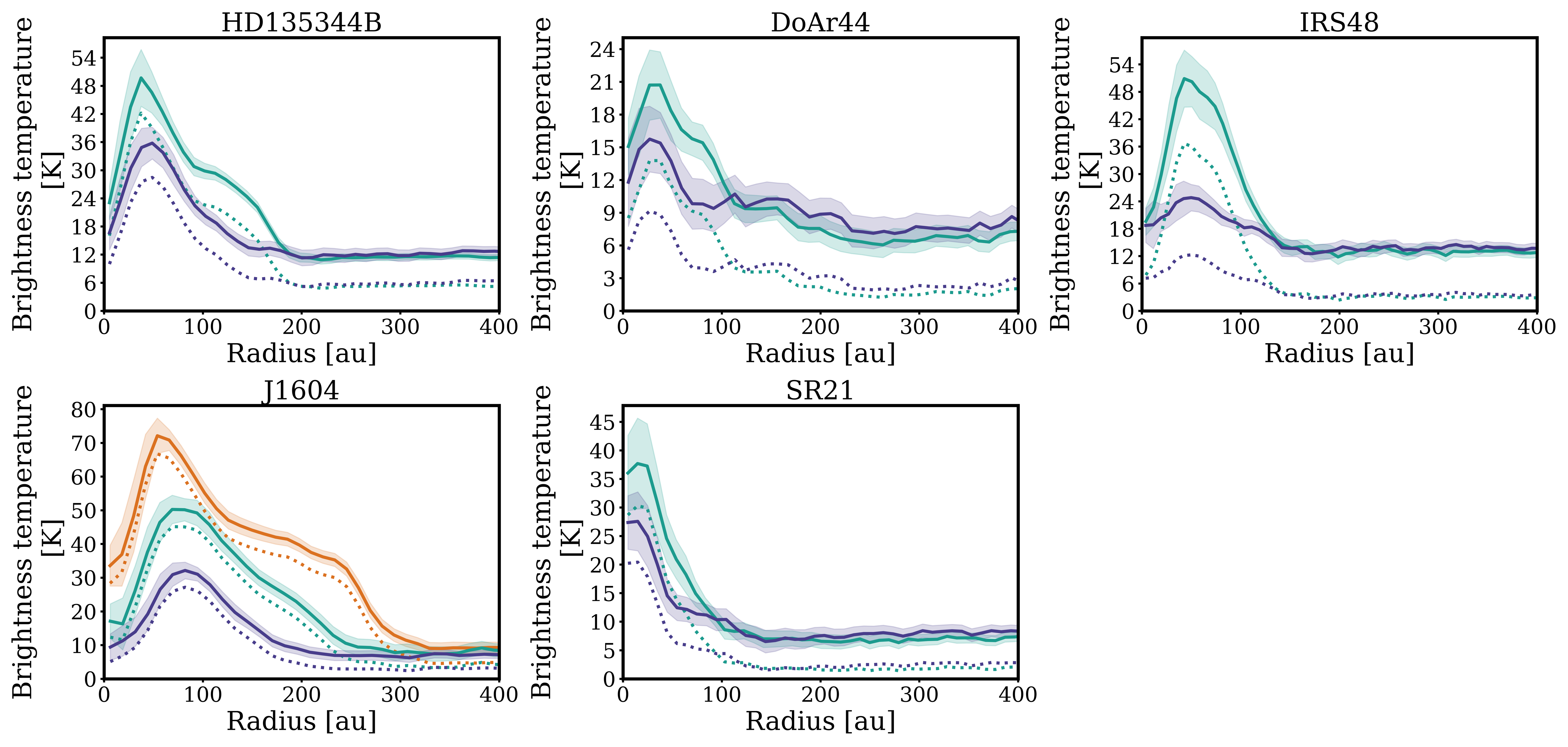}

      \caption{Deprojected azimuthally averaged radial profiles of the brightness temperature of the CO isotopologue emission. The solid lines show the brightness temperature calculated using the Planck function, while the dotted lines how the Rayleigh-Jeans approximation. $^{12}$CO ($J$=2-1, orange), $^{13}$CO ($J$=2-1, green), and C$^{18}$O ($J$=2-1, purple) are shown. \textit{Top row:} Disks analyzed in this work. \textit{Middle and bottom row:} Disks analyzed by \cite{2016vanderMarel}. The same profiles for HD169142 and LkCa 15 are reported by \cite{2022Leemker}.}
    \label{fig:Tb}  
\end{figure*}

For all the three disks in the sample, the top row in Figure \ref{fig:Tb} highlights the difference between the brightness temperature profiles estimated using the full Planck curve (solid lines) and the Rayleigh-Jeans approximation (dotted lines). The Rayleigh-Jeans approximation assumes that the emission is optically thick and spatially and spectrally resolved; otherwise, the brightness temperature is a lower limit to the actual kinetic temperature. Moreover, the approximation is valid when $h \nu \ll k T$, which results in $T \gg 11$ K for all CO isotopologues analyzed in this work ($^{12}$CO, $^{13}$CO, and C$^{18}$O $J$=2-1). This condition is fulfilled in only for the $^{12}$CO and $^{13}$CO line. For the C$^{18}$O line, the condition is marginally satisfied only in J1633. In the two other cases, since the Rayleigh-Jeans approximation does not apply, the estimate of $T_b$ is underestimated, and the full Planck function needs to be assumed. 

A direct comparison of the brightness temperatures shows that the three disks have a similar temperature profile in their emitting layers.
The brightness temperature of the  $^{12}$CO ($J$=2-1) is the highest and thus traces the warmer layer in the disks. Since the $^{12}$CO ($J$=2-1) line becomes optically thick high up in the disk, it traces the upper layer in the disk, close to the surface. The $^{13}$CO and C$^{18}$O are optically thinner and trace cooler layers, deeper down in the disks and closer to the mid-plane. In particular, $^{13}$CO is expected to be optically thick at the peak of the brightness temperature profile. This condition is reached at 30, 24, and 28 au, with a $T_b$ of 21, 32, and 28 K in Sz 111, J1608, and J1633, respectively. Since the $^{13}$CO brightness temperature profiles are calculated using the peak flux of the $^{13}$CO line, the location of the $^{13}$CO temperature peak $R_T$ does not necessarily correspond to the location of the $^{13}$CO intensity peak $R_\text{peak}$. Table \ref{tab:tpeak} reports the peak locations and the peak temperatures of the $^{13}$CO temperature profile for these disks, while Table \ref{tab:raddii_tab_mP} reports the location of the $^{13}$CO intensity peak.

\subsection{Inner dust disks}
\label{innerdustdisks}

Continuum emission is often detected within the cavities of transition disks \citep{2020FrancisAndVDMarel}. However, no such emission is detected in the disks analyzed in this work. Considering the rms levels as upper limits (Table \ref{tab:rms_beam_cont}), we can set an upper limit on the inner disk dust mass using the standard derivation of the dust mass from millimeter fluxes (see Section 3.2 and Equation (12) in \citealp{2020FrancisAndVDMarel}):

\begin{equation}\label{massEqTT}
    M_{\rm dust} = \frac{F_\nu d^2 \cos i}{\kappa_\nu\left( \frac{2 \nu^2 k_B}{c^2} \sqrt[4]{\frac{\phi L_*}{8 \pi \sigma_B}} \right)} 
\frac{(R_{\rm out} - R_{\rm in})}{2 \left(R_{\rm out}^{1/2} - R_{\rm in}^{1/2}\right)}.
\end{equation}

Here we assume three times the rms as a limit on the flux density $F_\nu$, $\nu =225$ GHz as the observing frequency, and an inner disk with outer radius $R_{\rm out}$ given by half of the beam size ($R_{\rm out} = \sqrt{\theta_\text{maj}\theta_\text{min}} / 2$). This equation assumes optically thin emission and the black-body approximation (i.e., ignoring scattering effects). $L_*$ is the stellar luminosity, $i$ is the disk inclination, $d$ is the distance of the disk, and $\sigma_B$ is the Stefan–Boltzmann constant. We adopt standard assumptions for the other parameters (see \citealp{2020FrancisAndVDMarel}): the inner radius of the inner disk is take as the sublimation radius, $R_{\rm in} = 0.07 \sqrt{L_*[L_\odot]}$ \citep{2001Dullemond}; the flaring angle $\phi$ of the disks is assumed to be $0.02$; the dust opacity $\kappa_\nu$ is taken as 10 cm$^{2}$ g$^{-1}$ at 1000 GHz, with an opacity power-law index of $\beta=1.0$ \citep{1991BeckwithANDSargent}. Table \ref{tab:sigma} reports the assumed $R_{\rm out}$ and the corresponding $M_{\rm dust}$.
The resulting upper limits on the inner disk dust masses are $<$0.003 $M_{\mathTerra}$, $<$0.02 $M_{\mathTerra}$, and $<$0.0006 $M_{\mathTerra}$ for Sz 111, J1633, and J1608, respectively.

The presence of an inner dust disk is also not expected based on the Spectral Energy Distribution (SED) of these three disks: in all disks, the SED shows no infrared excess until 24 $\mu$m \citep{2016vanderMarel-spitzer, 2018vanderMarel}, suggesting a lack of hot dust grains close to the star.
Nevertheless, the stellar accretion rates suggest that gas is still flowing through the gap and is present close to the host star. Using the relation described in \citet{2014Manara} between the accretion rate and the gas surface density, we estimate the expected gas surface densities at 1 au $\Sigma_\text{gas}(r=1 \rm au)$:

\begin{equation}\label{eq:sigma_gas}
    \Sigma_\text{gas}(r=1 \text{au}) \sim \frac{2 m_p}{3 \pi \alpha k_B T(r=R_{\rm out})} \dot M_\text{acc} \sqrt{\frac{G M_*}{R_{\rm out}^3}},
\end{equation}
where $m_p$ is the mass of the proton, $\alpha$ is the  turbulent viscosity parameter, and $T(r)$ is the midplane temperature profile (see below, Equation \ref{eq:generaleTemperature}).
Considering the observed accretion rates and assuming $\alpha$=10$^{-3}$, the estimated gas surface densities at 1 au are $\sim 1.0 \cdot10^2$, $\sim 0.2\cdot10^2$ and $\sim 1.9\cdot10^2$ g cm$^{-2}$ for Sz111, J1633 and J1608, respectively (see Table \ref{tab:sigma}). 
If an MHD wind drives the accretion, the viscosity close to the star would be much higher, with $\alpha>1$ \citep{2022Martel}, and the gas surface densities based on the accretion rate would be closer to $10^4-10^5$ g cm$^{-2}$.

\section{Discussion}\label{discussionp4}

\subsection{$^{13}$CO brightness temperature in transition disks}

Since the $^{13}$CO is expected to be optically thick at its peak, its brightness temperature at this location can be used as a diagnostic of the gas temperature to compare our sample to other transition disks. In contrast, $^{12}$CO emission typically originates from higher, more optically thick disk surface layers, making its brightness temperature less representative of the disk temperature structure.

The host stars in Sz 111, J1608, and J1633 are late-type stars, with spectral types M or K. Their disks are thus expected to be cooler than those around younger and more luminous stars. 
We compared our sample to seven other transition disks, reported in \cite{2016vanderMarel},  \cite{2017ADong}, and \cite{2022Leemker}: HD135344B, SR 21, DoAr 44, IRS 48, J160421.7-213028 (hereafter J1604), LkCa15, and HD169142. The first four disks were observed in Band 7, targeting the $J=3-2$ rather than the $J=2-1$ CO isotopologues. Although both the 2--1 and 6--5 transitions are presented in \citet{2022Leemker}, for consistency with our data, for LkCa15 and HD169142, we analyze only the 2--1 emission. DoAr 44, J1604, and LkCa15 are K-type stars, while SR21, HD169142, HD135344B, and IRS 48 are early-type stars with luminosity $> 7$ L$_\odot$, and therefore their disks are expected to be warmer than those in our sample.

\begin{table*}[htbp]
\centering
\small
\caption{ $^{13}$CO brightness temperature and radii}
\label{tab:tpeak}
\begin{tabular}{ccccccccc}
\hline\hline 
Target   & SpT & inc & $L_*$& $T_b$ & $R_\text{T}$ & $T_\text{mid}$ & Ratio & Reference \\
 & & [deg] & [L$_\odot$] & [K] &  [au] &   \\
\hline 

Sz 111 & M1.0 & 40 & 0.21 & 21 & 30 & 18 & 1.2 & This work\\

J1633 & K5.0 & 77 & 1.0 & 28 & 28  & 27 & 1.0 & This work\\

J1608 & K2.0 & 107 & 1.9 & 32 & 24 & 27 & 1.2 & This work\\

\hline

HD135344B &  F4 & 16 & 6.7 & 50  &   38 & 39 & 1.3 & 1 \\
DoAr 44 & K3 & 20 & 1.9 & 21 & 35 & 29 & 0.72 &  1 \\
IRS 48 & A0 & 50 & 17.8 & 51 & 43  & 39 & 1.3  & 1 \\
SR 21 & G3 & 16 & 11.0 & 38 & -  & - & -& 1 \\ 
J1604 & K2 & 6 & 0.7 & 50 & 65  & 14 & 3.6 & 2\\

LkCa15 & K3 & 50 & 1.1 &18 & 80  & 15 & 1.2 &  3 \\
HD169142 & F1 & 13 & 10 & 120 & 20 & 47 & 2.6 & 3  \\
\hline

\hline

\hline
\end{tabular}
\tablefoot{ The columns are the spectral type, the stellar luminosity $L_*$, the inclination, the brightness temperature  $T_b$ at the peak of the $^{13}$CO emission, the location of the $^{13}$CO temperature peak, $R_\text{T}$, the expected temperature $T_\text{mid}$ at $R_\text{T}$, and the ratio between $T_b$ and $T_\text{mid}$. See text for more details. References: 1) \cite{2016vanderMarel}, 2) \cite{2017ADong}, 3) \cite{2022Leemker}.}

\end{table*}

\begin{figure}
   \centering
   \includegraphics[width=0.5\textwidth]{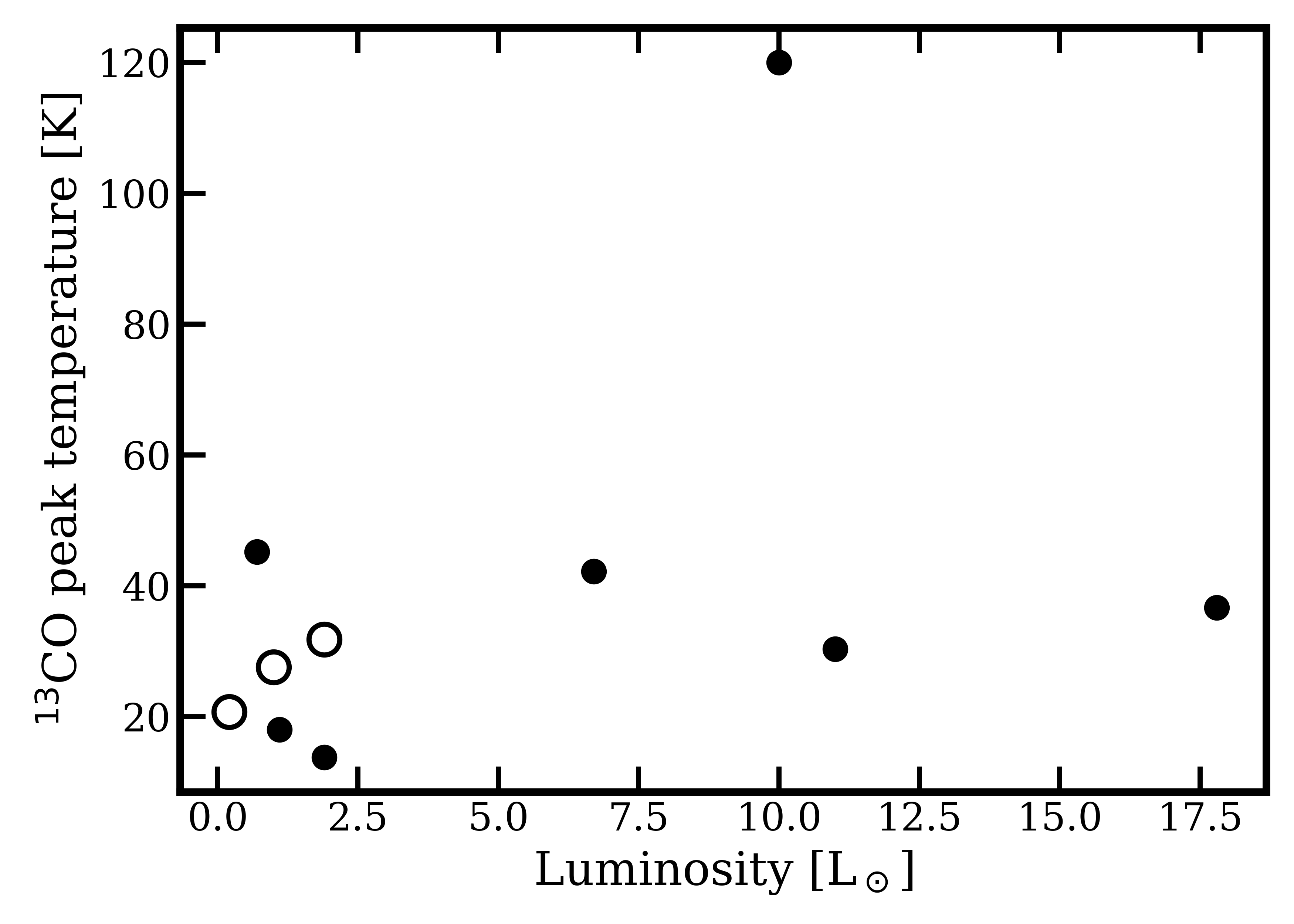}
   
     \caption{Brightness temperature at the peak of the $^{13}$CO emission as a function of the luminosity of the central star. Sz 111, J1633, and J1608 are shown with circles, while the other transition disks are shown with dots (see also Table \ref{tab:tpeak}).}
    \label{fig:temVSlum}  
\end{figure}

The middle and bottom rows in Figure \ref{fig:Tb} show the brightness temperature profiles of HD135344B, SR 21, DoAr 44, IRS 48, and J1604, derived from the data presented in \cite{2016vanderMarel} and \cite{2017ADong}. For LkCa15 and HD169142, we directly use the brightness temperature profiles of the $^{13}$CO $J=2-1$ line derived in Figure 4 of \citet{2022Leemker}. Table \ref{tab:tpeak} reports the peak locations $R_T$ and temperature peak of the $^{13}$CO line for these disks.
As expected from their stellar type, LkCa15, and DoAr 44 exhibit peak $^{13}$CO brightness temperature of $\sim 20$ K, comparable to that observed in Sz 111. Interestingly, J1604 shows a peak temperature of $\sim 50$ K, about twice as warm as J1608 and J1633, despite all three systems being K-type. 
SR21 and IRS 48, whose central stars are G3 and A0 types, show temperatures of 38 and 51 K, respectively. Although slightly higher, the value of the peak $^{13}$CO brightness temperature of SR21 is comparable to those of J1608 and J1633, despite the latter being K-type star systems. Notably, SR21 is the only system in which the $^{13}$CO does not show a cavity and peaks at the stellar location. CO rovibrational emission has indicated the presence of a small gas cavity of 7 au \citep{2008Pontoppidan}, which was indeed seen (but not quantified) in high-resolution ALMA CO data presented in \citet{Yang2023}.
HD135344B and HD169142 are F4 and F0 type stars and show a peak temperature of 50 and 120 K, as expected, considerably higher than those of our sample.
Figure \ref{fig:temVSlum} shows the temperature peak of the $^{13}$CO line as a function of the stellar luminosity. No significant trend or differences are found in the disk temperature of disks around more or less luminous stars. 
This suggests that the gas temperature is not only regulated by the stellar luminosity, but also by additional parameters related to the detailed disk structure and chemistry, such us the presence of an inner disk, the fraction of large grains, or the depth of the gas cavity (e.g., \citealp{2022Leemker}).

The brightness temperature profiles (Figure \ref{fig:Tb}) are calculated using the peak flux of the CO lines (i.e., the eighth moment maps), under the assumption that the spatial resolution is sufficient to resolve the $^{13}$CO peak. At lower spatial resolution, the inferred $T_b$ is expected to be underestimated due to beam smearing. \cite{2022Leemker} report that the spectral and spatial resolution of the data can affect the estimate of the brightness temperature by $\sim 10-30\%$. New ALMA observations of the CO isotopologues, with a uniform spectral and spatial resolution for all disks, are thus crucial for performing a detailed comparison of the brightness temperature profiles. 

As a comparison, we calculated the expected midplane temperature at the location $R_T$ of the peak of the $^{13}$CO temperature profile, assuming the standard simplified expression for a full flared disk (assumed flaring angle $\phi =0.02$), passively heated in radiative equilibrium (e.g., \citealp{1997Chianf&Goldreich}, \citealp{2001Dullemond}):

\begin{equation}\label{eq:generaleTemperature}
    T_\text{mid}(r) = \left( \frac{\phi L_*}{8 \pi \sigma_B r^2} \right)^{1/4},
\end{equation}
calculated at $r = R_T$. Table \ref{tab:tpeak} reports the expected midplane temperature $T_\text{mid}$ and the ratio between the observed peak of the $^{13}$CO temperature profile and this value.  This ratio is typically in the range $0.7-1.3$, except for J1604 and HD169142, where it reaches $\sim 3$.
The high ratios of $\sim 3$ in J1604 and HD169142 indicate a higher surface temperature compared to the other disks, whereas low ratios (average value $\sim 1.1$) suggest optically thinner emission, as the gas temperature is not expected to fall below the dust temperature due to gas-dust coupling. 

The peak brightness temperature of the $^{13}$CO $J=2-1$ line for our transition disks, as well as those from \cite{2016vanderMarel} and \cite{2017ADong}, is consistent with the $^{13}$CO $J=2-1$ and $J=3-2$ temperatures found in a sample full disks or gapped disk around Herbig and T Tauri stars ($L_* \sim 1-20$ L$_\odot$), where peak temperatures of 20-40 K are found \citep{2023Paneque-Carreno}. The only exception is HD169142, which shows a much higher peak temperature of 120 K.

These comparisons between transition disks around late- and early-type stars, as well as between transition and full disks, do not reveal any significant differences when using the $^{13}$CO to derive the brightness temperature close to the midplane. This suggests that caution is required when deriving temperatures, as optical depth effects must be taken into account. In particular, disk inclination can impact the optical depth along the line-of-sight optical and the vertical emitting layer traced by the CO emission, affecting the observed brightness temperature. Dedicated studies of individual targets through detailed thermochemical modeling (e.g., with DALI, \citealp{2013Bruderer}) or multi-transition analysis \citep{2022Leemker} are needed to constrain the properties of the gas and dust cavities.

\begin{table}[htbp]
\centering
\small
\caption{Radii from the $^{13}$CO intensity profile}
\label{tab:raddii_tab_mP}
\begin{tabular}{cccccc}
\hline\hline 
Target  & $R_\text{peak}$ & $R_\text{CO}$ & $R_\text{gap}$ & $M_p$ & Reference \\
  &  [au] &  [au] &  [au] & [$M_\text{jup}$] &\\
\hline 

Sz 111  & 23 & 17 & 13 & 87 & This work\\
J1633  & 22  & 16 & 12 & 61 & This work\\
J1608  & 30 &  22 & 1 & 15 & This work\\

\hline

HD135344B &   27 &  19 & 15 & 280 & 1 \\
DoAr 44  & 25 & 18 & 14 & 227 & 1 \\
IRS 48 & 43 & 31 &  24 & 77 & 1 \\
SR 21  & - & - & - & - & 1 \\
J1604 & 65 & 46 & 36 & 17 & 2 \\

HD169142  & 20 & 14 & 11 & 24 & 3 \\
\hline

HD142527 &   118 &  84 & 65 & 52 & 4 \\
AB Aur &  117 & 84 & 64 & 34 & 4 \\
MWC 758 &  40 & 29 & 22  & 10 & 4 \\
CQ Tau &  23 & 16 & 13 & 869 & 4 \\
LKCa 15 &  45 & 32 & 25 & 102 & 3, 4 \\
PDS 70 & 41 & 29 & 23 & 192 & 4 \\
Sz 91 &  52 & 37 & 29 & 185 & 4 \\

\hline

\hline
\end{tabular}
\tablefoot{ The columns are the location of the $^{13}$CO peak, $R_\text{peak}$; gap edge $R_\text{CO}$ of the $^{13}$CO emission; gap radii $R_\text{gap}$ of the $^{13}$CO emission. The mass of the possible companion $M_p$ carving the cavity is estimated following \cite{2018Facchini}. $M_p>15$ M$_\text{jup}$ are outside the range studied by \cite{2018Facchini}. See text for more information. References: 1) \cite{2016vanderMarel}, 2) \cite{2017ADong}, 3) \cite{2022Leemker}, 4) \cite{2021vdMarel}. }

\end{table}

\subsection{Gas to dust ratio in the inner disk}
\label{gastodust}

The deep upper limits on the inner dust disk mass of at most $<0.02 M_{\mathTerra}$ of our targets (see Section \ref{innerdustdisks}) are significantly lower than typical values of such inner disks in other transition disks \citep{2020FrancisAndVDMarel}. However, recent work has demonstrated that the detected central continuum emission in transition disks is more likely dominated by free-free emission, associated with the presence of MHD winds or jets connected with the accretion rate onto the star \citep{2024Rota}.
\citet{2025Garufi} show that the free-free emission contributes on average $65\%$ of the total flux at 2 cm, while its contribution is only $2\%$ at 1.3 mm. Using ALMA Band 6 and 7 observations and assuming that the emission is dominated by thermal dust emission, the inner dust disk masses range from $0.003-0.65$ $M_{\mathTerra}$ in their sample of transition disks (see Table 5 in \cite{2024Rota}), comparable to our upper limits. This suggests that the amount of dust in the inner disk in transition disks is generally very low.

Starting from these upper limits on the inner dust mass, we can estimate the expected dust surface density $\Sigma_\text{dust}$ at 1 au, using the same approach described in \citet{2020FrancisAndVDMarel} (see Equation 11 and Section 3.3): 
\begin{equation}
    \Sigma_\text{dust} (r_c=1 \text{ au}) \frac{M_\text{dust}}{2 \pi (R_\text{out} - R_\text{in})} \frac{1}{r_c},
\end{equation}
with $r_c = 1$ au and $R_{\rm out} = \sqrt{\theta_\text{maj}\theta_\text{min}} / 2$ (see assumption for $M_\text{dust}$ in Section \ref{innerdustdisks}).
Considering the gas surface densities based on the stellar accretion rates (see Equation \ref{eq:sigma_gas}), it is possible to derive the gas-to-dust ratio in the inner disk at 1 au separation. As reported in Table \ref{tab:sigma}, the gas-to-dust ratios vary over one order of magnitude in Sz 111, J1633, and J1608, while they vary over two orders of magnitude in the sample of transition disks studied by \citet{2024Rota}. No evident differences between the two samples are found. 

\begin{figure}
   \centering
   \includegraphics[width=0.5\textwidth]{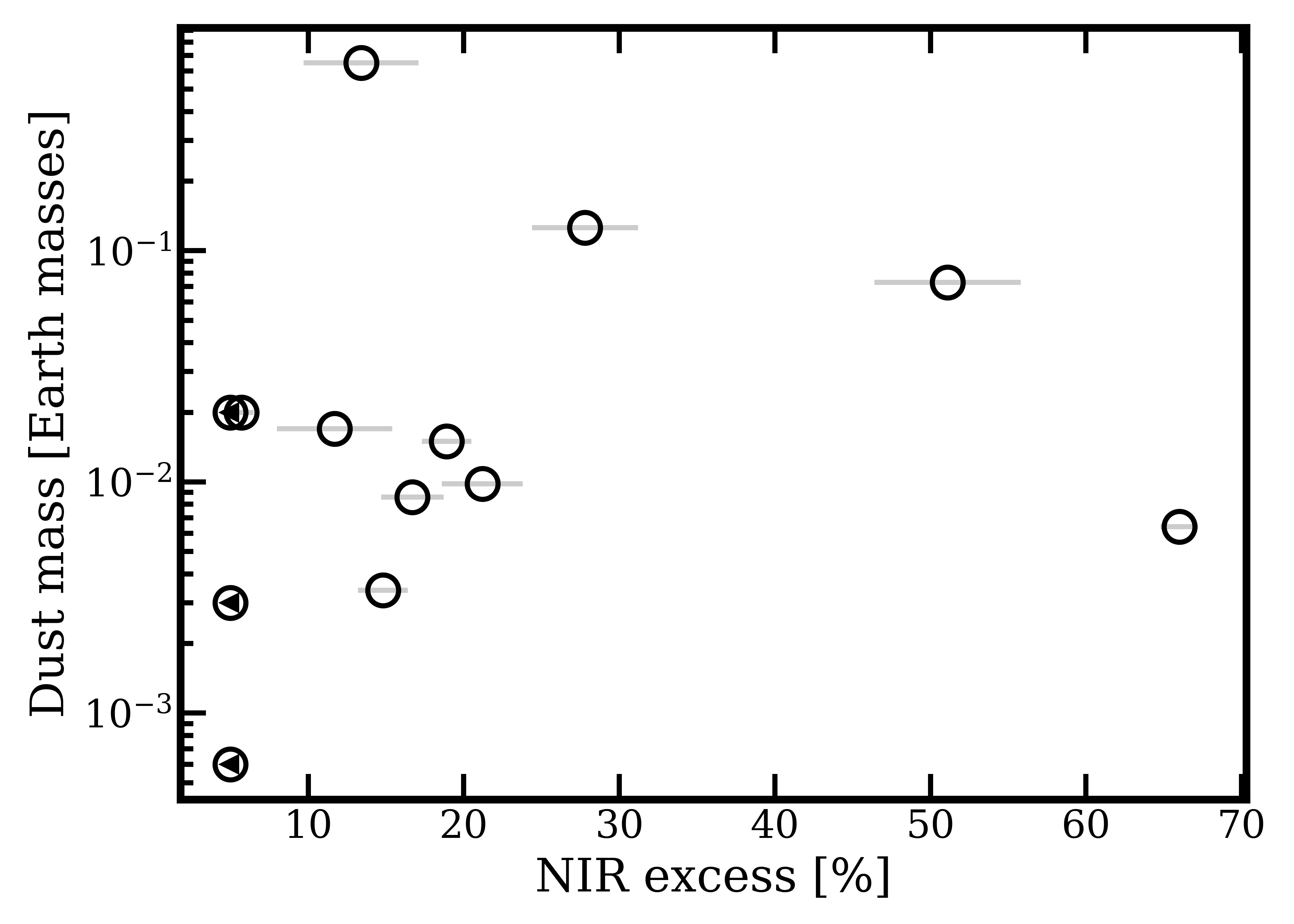}
   
      \caption{Inner disk dust masses as a function of the NIR excess (reported by \citealp{2020FrancisAndVDMarel}) for our transition disks and the sample of disks analyzed in \citet{2024Rota}. Upper limits on the NIR excess are shown with triangles.}
    \label{fig:NIR}  
\end{figure}

Figure \ref{fig:NIR} shows the inner disk dust masses as a function of the NIR excess (reported by \citealp{2020FrancisAndVDMarel}) for our transition disks and the sample of disks analyzed in \citet{2024Rota}. No correlation between these quantities is found even after taking into account the free-free emission. A lack of correlation was previously reported for a large sample of transition disks by \cite{2020FrancisAndVDMarel}, but their dust masses assumed an outer radius of the inner disk of 5 au and may have been contaminated with free-free emission. In this work, we assume as an upper limit on $R_{\rm out}$ half of the beam size and higher resolution images. A different assumption on $R_{\rm out}$ changes the dust surface density at 1 au, and consequently, the inner disk dust mass. The new estimates are thus more self-consistent and show that very little dust is left in the inner disks.

\begin{table*}[htbp]
\centering
\small
\caption{Gas and dust surface density at 1 au and near-infrared excess}
\label{tab:sigma}
\begin{tabular}{ccccccc}
\hline\hline 
Target  & $R_\text{out}$ & $M_\text{dust}$  & $\Sigma_\text{dust}$ & $\Sigma_\text{gas}$ & $\Sigma_\text{gas}/\Sigma_\text{dust}$  & NIR excess \\
  & [au] & [M$_\mathTerra$] & [g cm$^{-2}$] &  [g cm$^{-2}$] &  & [\%]\\
\hline

Sz111 & $<2.5$ & $<0.003$  & $<0.53\times 10^{-2} $ & $1.0\times 10^{2} $  & $>1.9\times 10^{4}$ & $<5$ \\
J1633 & $<2.4$ &  $<0.02$ & $<0.38\times 10^{-2} $  & $0.2\times 10^{2} $  & $>0.6\times 10^{4}$ & $<5$ \\
J1608 & $<2.0$ & $<0.0006$ & $<0.15\times 10^{-2} $  & $1.0\times 10^{2} $  & $>12\times 10^{4}$ & $<5$ \\
 \hline
AB Aur & $<2.4$ & $<0.015$ & $<3.23\times 10^{-2} $  & $1130\times 10^{2} $  & $>350\times 10^{4}$ & $18.9\pm1.6$ \\
GGTau & $<12.1$ & $<0.65$ & $<23.0\times 10^{-2} $  & $97\times 10^{2} $  & $>4.2\times 10^{4}$ & $13.4\pm3.7$ \\
HD100453 & $<1.7$ & $<0.0086$ & $<2.40\times 10^{-2} $  & $32\times 10^{2} $  & $>13\times 10^{4}$ & $16.7\pm2.0$ \\
HD100546 & $<1.1$ & $<0.020$ & $<11.20\times 10^{-2} $  & $430\times 10^{2} $  & $>38\times 10^{4}$ & $5.7\pm0.8$ \\
HD142527 & $<10.1$ & $<0.073$ & $<3.17\times 10^{-2} $  & $430\times 10^{2} $  & $>140\times 10^{4}$ & $51.1\pm4.7$ \\
HD169142 & $<3.2$ & $<0.0064$ & $<0.95\times 10^{-2} $  & $140\times 10^{2} $  & $>150\times 10^{4}$ & $66\pm1.2$ \\
SR24S & $<2.2$ & $<0.017$ & $<3.44\times 10^{-2} $  & $180\times 10^{2} $  & $>52\times 10^{4}$ & $11.7\pm3.7$ \\
TCha & $<12.3$ & $<0.126$ & $<4.36\times 10^{-2} $  & $11\times 10^{2} $  & $>2.4\times 10^{4}$ & $27.8\pm3.4$ \\
HD135344B & $<14.3$ & $<0.0098$ & $<0.30\times 10^{-2} $  & $110\times 10^{2} $  & $>360\times 10^{4}$ & $21.2\pm2.6$ \\
MWC758 & $<2.8$ & $<0.0034$ & $<0.56\times 10^{-2} $  & $180\times 10^{2} $  & $>320\times 10^{4}$ & $14.8\pm1.6$ \\

\hline
\end{tabular}
\tablefoot{ The columns are the outer radius $R_\text{out}$ of the inner disk, estimated from the beam size; the inner disk dust mass $M_\text{dust}$; the gas and dust surface density at 1 au ($\Sigma_\text{gas}$ and $\Sigma_\text{dust}$, respectively), and their ratio. The last column reports the NIR excess from \citep{2020FrancisAndVDMarel}. See text for more details.
}
\end{table*}

\subsection{Companion masses}

The gas distribution inside the cavity of transition disks has often been shown to have a gradual drop in density (e.g., \citealp{2016vanderMarel}, \citealp{2017ADong}), consistent with disk clearing by a companion. The companion is expected to be located at the minimum of the gas density drop (i.e., at $R_\text{gap}$), with the dust typically trapped outside the outer edge of the cavity \citep{2012Pinilla}. The morphology of the cavity, traced both in (sub-)millimeter continuum emission and CO lines, can therefore be used to infer the mass of the planet that is expected to carve the disk.

\cite{2018Facchini} performed the post-processing of hydrodynamical simulations of disks hosting a single massive planet ($M_p \geq 1 M_\text{Jup}$), using ray-traced low-$J$ CO observations to derive an analytical relation between the cavity sizes and the planetary mass. They found a linear dependency in the log-log space, given by

\begin{equation}\label{eq:massss}
    \log_{10} \bigg( \frac{R_\text{mm}}{R_\text{CO}} \bigg) = \tilde{A} + \tilde{B} \log_{10} \bigg( \frac{M_p}{M_\text{jup}} \bigg),
\end{equation}
where $\tilde{A} = 0.101$ and $\tilde{B} = 0.134$. $R_\text{mm}$ is the radius of the sub-mm ring (see Table \ref{tab:fluxes_radii}), while $R_\text{CO}$ is defined as the location where the CO emission reaches 33\% of the flux difference between the bottom of the gap and maximum flux outside the gap. Since the resolution of our observation is insufficient to resolve the CO gap, $R_\text{CO}$ cannot be measured directly. Instead, we followed the same approach as \cite{2021vdMarel}, who used the location of the peak of the $^{13}$CO emission, $R_\text{peak}$, as a proxy. This peak location is approximately 1.4 times larger than $R_\text{CO}$. Furthermore, for planet mass between $5-15 M_\text{Jup}$, $R_\text{CO}$ can be related to $R_\text{gap}$, the minimum of the gas density drop, as $R_\text{CO} / R_\text{gap} = 1.3$ (\citealp{2018Facchini}, \citealp{2021vdMarel}). In this way, we can estimate the expected mass of the companion in the cavities. Table \ref{tab:raddii_tab_mP} reports the location of $R_\text{peak}$, $R_\text{CO}$, and $R_\text{gap}$, and the derived $M_p$. 

The companion masses estimated for Sz 111, J1608, and J1633 are in all cases $>15M_\text{Jup}$, and thus are in the brown dwarf/substellar regime.  
These masses are outside the parameter space explored by \cite{2018Facchini}, who restricted their study to $M_p < 15 M_\text{Jup}$ planets. Therefore, their scaling relation (Equation \ref{eq:massss}) cannot be reliably extrapolated to higher masses. Furthermore, \cite{2018Facchini} neglect eccentricity effects that are expected at mass ratios above $q \gtrsim 0.003$ (i.e., $\gtrsim$3 $M_{Jup}$ for Solar type stars), where Lindblad resonances can induce gap eccentricity and disk back-reaction can further amplify the companion’s orbital eccentricity (\citealp{2006Kley&Dirksen}; \citealp{2006DAlessio}; \citealp{2018Ragusa}), making the derivation of companion masses from the CO cavity using this methodology less reliable.
A similar result is reported by \cite{2021vdMarel} for a larger sample of transition disks. Table \ref{tab:raddii_tab_mP} summarizes the peak location and planetary mass estimated for their sample. Using high-contrast direct imaging, \cite{2021vdMarel} derived upper limits on potential companion masses, ruling out (sub)stellar companions with $M_p>50M_\text{Jup}$ at the gap location in most systems. This suggests that the observed disk structures can be explained by super-Jovian companions on eccentric orbits, but further modeling and deeper high-resolution imaging, such as will be possible with ELT-METIS, are needed to confirm this conclusion. 

\section{Conclusions}\label{conclusionsp4}

In this work, we investigated the dust and gas structures in three inclined ($i>40 \deg$) transition disks 
around late-type stars, 
using ALMA Band 6 observations of the dust continuum and CO $J$=2–1 isotopologues ($^{12}$CO,  $^{13}$CO, and  C$^{18}$O) at $\sim0\farcs02$ and $\sim0\farcs1$ resolution, respectively. 

\begin{enumerate}
    \item In all three disks, the continuum emission reveals a central cavity surrounded by a narrow, azimuthally symmetric dust ring. The dust cavities extend to $\sim 40$ au in radius, while the gas disks extend to $\sim 60-200$ au. 
    \item J1633 shows the narrowest ring, while the radial profiles of Sz 111 and J1608 present a shoulder at $\sim90$ au, suggesting the possible presence of a second unresolved gap.

\item The CO isotopologue lines are detected with a pattern compatible with Keplerian rotation, with no significant foreground absorption. Similar to the dust component, the CO isotopologues exhibit central cavities, with line emission detected within the dust cavities. 
\item The $^{12}$CO emission peaks closer to the star than both the $^{13}$CO and C$^{18}$O, likely reflecting the effect of different optical depths according to the CO isotopologues' different abundances. 
\item The brightness temperature of the $^{12}$CO is the highest and traces the warmer, upper layer in the disks. The $^{13}$CO and C$^{18}$O are optically thinner and trace cooler layers, deeper down in the disks and closer to the mid-plane. In particular, the $^{13}$CO line, expected to be optically thick at the peak of the brightness temperature profile, shows peak temperatures of $\sim 15-30$ K at $\sim 25-30$ au.

 \item As expected from the SED of these disks, showing no infrared excess up to $24\mu$m, no inner disks are detected down to the rms level. If an inner disk of dust is present, this translates into upper limits on the inner disk dust masses are 0.003 $M_{\mathTerra}$, 0.02 $M_{\mathTerra}$, and 0.0006 $M_{\mathTerra}$ for Sz 111, J1633, and J1608, respectively.
 On the other hand, the stellar accretion rates suggest that gas is still flowing through the gap.
 The gas-to-dust ratios at 1 au are thus $\sim 10^5$.

 \item No significant differences in the brightness temperature or in the gas-to-dust ratio are found between our disks and other transition disks or full disks, contrary to expectations. This suggests that parameters other than luminosity play a role in setting the gas temperature structure, requiring individual-target studies through thermochemical modeling or multi-transition analyses to constrain the properties of the dust and gas cavities.

 \item Assuming that the gradual drop of the gas distribution in the cavity and the dust ring are the effect of disk clearing by a companion, the location of the continuum peak and the CO radius can be used as a proxy to estimate the mass of the companion \citep{2018Facchini}. The observed disk structures could be explained by super-Jovian companions on eccentric orbits, but more detailed modeling is needed to confirm this conclusion.

 \end{enumerate}

\begin{acknowledgements}
The authors thank the referee for their useful comments that have contributed to improving the manuscript.
EvD acknowledges support from A-ERC grant 101019751 MOLDISK.
This paper makes use of the following ALMA data: 2018.1.00689.S.
ALMA is a partnership of ESO (representing its member states), NSF (USA) and NINS (Japan), together with NRC (Canada), MOST and ASIAA (Taiwan), and KASI (Republic of Korea), in cooperation with the Republic of Chile. The Joint ALMA Observatory is operated by ESO, AUI/NRAO and NAOJ. The PI acknowledges assistance from Allegro, the European ALMA Regional Center node in the Netherlands.     
\end{acknowledgements}

\bibliography{bibliography.bib}

\appendix

\onecolumn
\section{Radii estimate}

This appendix presents the cumulative flux curves used to estimate the disk radius and total flux in the dust and gas components. 

\begin{figure*}
   \centering
    \includegraphics[width=0.95\textwidth]{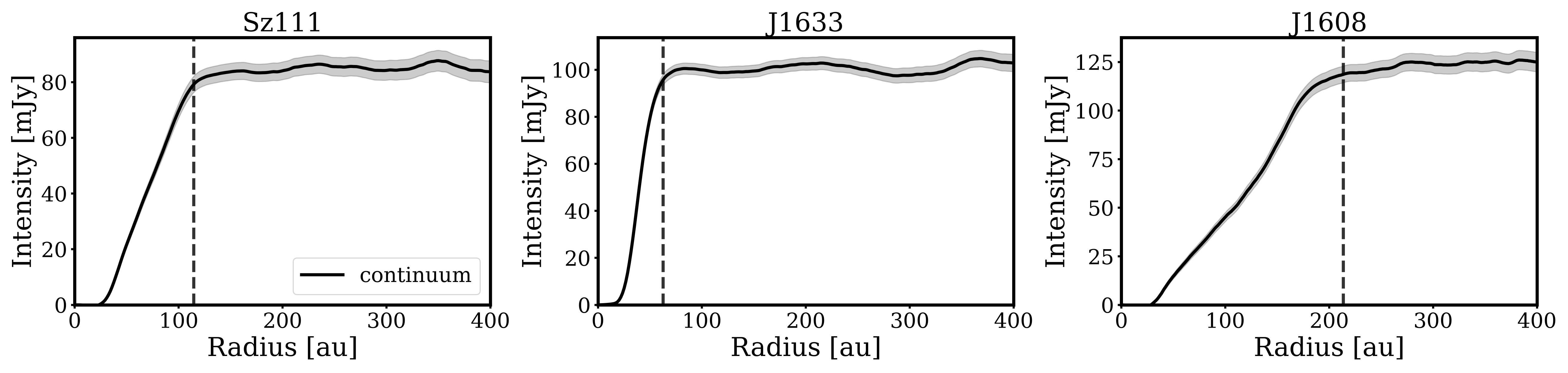}
    \includegraphics[width=0.95\textwidth]{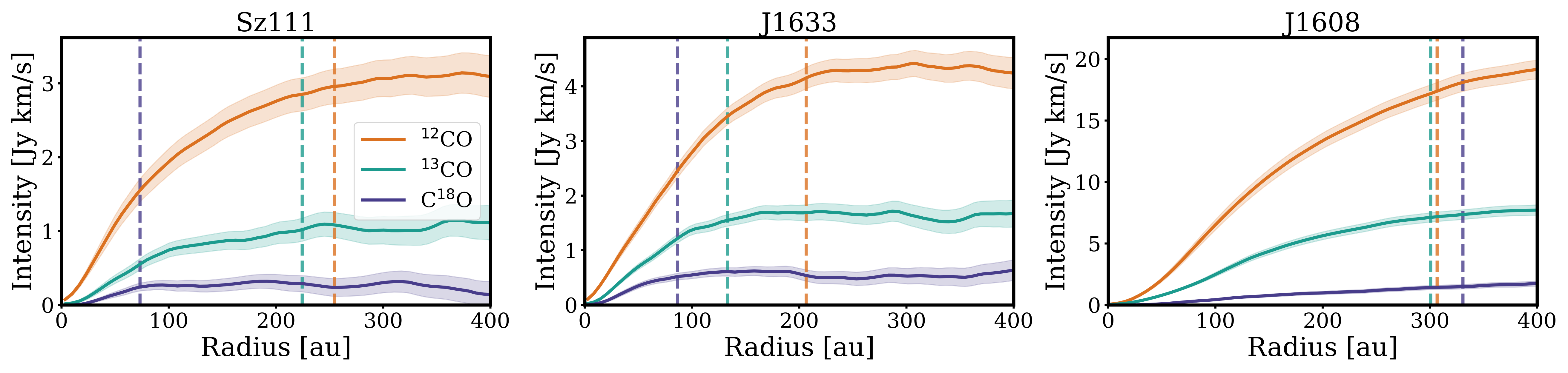}

      \caption{Cumulative flux curves as a function of the disk radius for the continuum emission (\textit{top}) and for the CO isotopologues (\textit{bottom}) emission, $^{12}$CO ($J$=2-1, orange), $^{13}$CO ($J$=2-1, green), and C$^{18}$O ($J$=2-1, purple). The dashed vertical lines show the estimated gas and dust disk radii, defined as the radii containing 95\% of the total fluxes.}
    \label{fig:cumulatives}  
\end{figure*}

\end{document}